\documentclass[11pt,twoside,a4paper]{article}

\usepackage{natbib}
\usepackage{graphicx}

\usepackage{url}
\usepackage{scrextend}
\usepackage{amsmath}
\usepackage{amssymb}
\usepackage{epstopdf}
\usepackage{mathrsfs}
\usepackage{float}
\usepackage{ragged2e}

\mathchardef\mhyphen="2D
\floatstyle{ruled}
\newfloat{conversation}{thp}{lop}
\floatname{conversation}{Conversation}
\floatstyle{boxed}
\restylefloat{conversation}
\newenvironment{qFont}{\fontfamily{lmr}\selectfont \footnotesize}{\par}

\title{Artificial Intelligence is stupid and causal reasoning wont fix it}
\author{J. Mark Bishop$^1$}
\date{%
    \small
    $^1$ TCIDA, Goldsmiths, University of London, London, UK\\
    Email: m.bishop@gold.ac.uk\\
    \normalsize
}

\begin{document}
\bibliographystyle{agu}

\maketitle

\section*{Abstract}
Artificial Neural Networks have reached `Grandmaster' and even `super-human' performance' across a variety of games, from those involving perfect-information, such as Go \citep{Silver_2016}; to those involving imperfect-information, such as `Starcraft' \citep{Vinyals_2019}. Such technological developments from AI-labs have ushered concomitant applications across the world of business, where an `AI' brand-tag is fast becoming ubiquitous. A corollary of such widespread commercial deployment is that when AI gets things wrong - an autonomous vehicle crashes; a chatbot exhibits `racist' behaviour; automated credit-scoring processes `discriminate' on gender etc. -  there are often significant financial, legal and brand consequences, and the incident becomes major news. 

As Judea Pearl sees it, the underlying reason for such mistakes is that ``\emph{... all the impressive achievements of deep learning amount to just curve fitting}''. The key, Pearl suggests \citep{Pearl_2018a}, is to replace `reasoning by association' with `causal reasoning' - the ability to infer causes from observed phenomena. It is a point that was echoed by Gary Marcus and Ernest Davis in a recent piece for the New York Times: ``\emph{we need to stop building computer systems that merely get better and better at detecting statistical patterns in data sets -- often using an approach known as ``Deep Learning'' -- and start building computer systems that from the moment of their assembly innately grasp three basic concepts: time, space and \textbf{causality}}'' \citep{Marcus_2019}. 

In this paper, foregrounding what in 1949 Gilbert Ryle termed `a category mistake' \citep{Ryle_1949}, I will offer an alternative explanation for AI errors; it is not so much that AI machinery cannot `grasp' causality, but that AI machinery (qua computation) cannot understand anything at all.\\

\small \textbf{Keywords}: Cognitive Science, Artificial Intelligence, Artificial Neural Networks, Causal Cognition, Chinese Room Argument, Dancing with Pixies.
\normalsize

\section{Making a mind}
For much of the twentieth century the dominant cognitive paradigm identified the mind with the brain; as the Nobel laureate Francis Crick eloquently summarised:

\begin{quote}
\begin{qFont}
\justify
``You, your joys and your sorrows, your memories and your ambitions, your sense of personal identity and free will, are in fact no more than the behaviour of a vast assembly of nerve cells and their associated molecules. As Lewis Carroll's Alice might have phrased, ``You're nothing but a pack of neurons''. This hypothesis is so alien to the ideas of most people today that it can truly be called astonishing'' \citet{Crick_1994}.
\end{qFont}
\end{quote}

Motivation for the belief that a computational simulation of the mind is possible stemmed initially from the work of \citet{Turing_1936} and \citet{Church_1936} and the `Church-Turing hypothesis'; in Turing's formulation, every `function which would naturally be regarded as computable' can be computed by the `Universal Turing Machine'. If computers can adequately model the brain then, theory goes, it ought to be possible to \emph{program} them to act like minds. And consequently, in the latter part of the twentieth century, Crick's ``Astonishing Hypothesis'' helped fuel an explosion of interest in connectionism: both high-fidelity simulations of the brain (computational neuroscience; theoretical neurobiology) and looser - merely `neural inspired' - analogues (cf. Artificial Neural Networks; Multi-Layer Perceptrons; `Deep Learning' systems).

But the fundamental question that Crick's hypothesis raises is, of course, this: if we ever succeed in fully instantiating a \emph{sufficiently accurate} simulation of the brain on a digital computer, will we also have fully instantiated a digital [computational] mind, with all the human mind's causal power of teleology, understanding and reasoning; will AI finally have succeeded in delivering `Strong AI'\footnote{Strong AI, a term coined by \citet{Searle_1980} in the `Chinese Room Argument' (CRA), entails that, ``\emph{... the computer is not merely a tool in the study of the mind; rather, the appropriately programmed computer really is a mind, in the sense that computers given the right programs can be literally said to understand and have other cognitive states}'', which Searle contrasted with ``Weak AI'' wherein ``\emph{... the principal value of the computer in the study of the mind is that it gives us a very powerful tool.}'' I.e. Weak AI focusses on epistemic issues relating to engineering a simulation of [human] intelligent behaviour whereas Strong AI, in seeking to engineer a computational system with all the causal power of a mind, focusses on the ontological.}

Of course, \emph{if} Strong AI is possible, accelerating progress in its underpinning technologies\footnote{Cf. ``[A]mplifiers for intelligence - devices that supplied with a little intelligence will emit a lot'', \citep{Ashby_1956}.} - entailed both by the use of AI systems to design ever more sophisticated AIs and the continued doubling of raw computational power every two years\footnote{Cf. Moore's `law' :- the observation that the number of transistors in a dense integrated circuit approximately doubles every two years.} - will eventually cause a runaway effect whereby the Artificial Intelligence will inexorably come to exceed human performance on all tasks\footnote{Conversely, as Francois Chollet, a senior engineer at Google and well known sceptic of the `Intelligence Explosion' scenario; trenchantly observed in 2017: ``\emph{The thing with recursive self-improvement in AI, is that if it were going to happen, it would already be happening. I.e. Auto Machine Learning systems would come up with increasingly better Auto Machine Learning systems, Genetic Programming would discover increasingly refined GP algorithms}'' and yet, as Chollet insists, ``\emph{no human, nor any intelligent entity that we know of, has ever designed anything smarter than itself}''.}; the so-called point of [technological] `singularity' ([in]famously predicted by Ray Kurzweil to occur as soon as 2045\footnote{\citet{Kurzweil_2005} ``set the date for the Singularity - \emph{representing a profound and disruptive transformation in human capability} - as 2045''.}). And, at the point this `singularity' occurs, so commentators like Kevin Warwick\footnote{In his 1997 book ``March of the Machines'' \citep{Warwick_1997} observed that there were already robots with the `\emph{brain power of an insect}'; soon, or so he predicted, there would be robots with the `\emph{brain power of a cat}', and soon after that there would be `\emph{machines as intelligent as humans}'. When this happens, Warwick darkly forewarned, the science-fiction nightmare of a `Terminator' machine could quickly become reality because such robots will rapidly, and inevitably, become more intelligent and superior in their practical skills than the humans who designed and constructed them.} and Stephen Hawking\footnote{In a television interview with Professor Stephen Hawking on December 2nd 2014, Rory Cellan-Jones asked how far engineers had come along the path towards creating Artificial Intelligence, to which Professor Hawking alarmingly replied, ``\emph{Once humans develop artificial intelligence it would take off on its own and redesign itself at an ever increasing rate. Humans, who are limited by slow biological evolution, couldn't compete, and would be superseded.}''} suggest, humanity will, effectively, have been ``superseded'' on the evolutionary ladder and be obliged to eke out its autumn days listening to `Industrial Metal' music and gardening; or, in some of Hollywood's even more dystopian dreams, cruelly subjugated (and/or exterminated) by `Terminator' machines.

In this paper, however, I will offer a few `critical reflections' on one of the central, albeit awkward, questions of AI: why is it that, over seven decades since Alan Turing first deployed an `effective method' to play chess in 1948, we have seen enormous strides in engineering particular machines to do clever things -- from driving a car to beating the best at Go -- but almost no progress in getting machines to genuinely understand; to seamlessly apply knowledge from one domain into another -- the so-called problem of `Artificial General Intelligence' (AGI); the skills that both Hollywood and the wider media really think of, and depict, as Artificial Intelligence?

\section{Neural Computing}
The earliest Cybernetic work in the burgeoning field of `neural computing' lay in various attempts to understand, model and emulate neurological function and learning in animal brains, the foundations of which were laid in 1943 by the neurophysiologist Warren McCulloch and the mathematician Walter Pitts \citep{McCulloch_1943}. 

Neural Computing defines a mode of problem solving based on `learning from experience' as opposed to classical, syntactically specified, `algorithmic' methods; at its core is ``\emph{the study of networks of `adaptable nodes' which, through a process of learning from task examples, store experiential knowledge and make it available for use}'' \citep{Aleksander_1995}. So construed, an `Artificial Neural Network' (ANN) is constructed merely by appropriately connecting a group of adaptable nodes (`artificial neurons').
\begin{itemize}
\item A \emph{single layer neural network} only has one layer of adaptable nodes between the input vector, $X$ and the output vector $O$, such that the output of each of the adaptable nodes defines one element of the network output vector $O$.
\item A \emph{multi-layer neural network} has one or more `hidden layers' of adaptable nodes between the input vector and the network output; in each of the network \emph{hidden layers}, the outputs of the adaptable nodes connect to one or more inputs of the nodes in subsequent layers and in the network \emph{output layer}, the output of each of the adaptable nodes defines one element of the network output vector $O$.
\item A \emph{recurrent neural network} is a network where the output of one or more nodes is fed-back to the input of other nodes in the architecture, such that the connections between nodes form a `directed graph along a temporal sequence', so enabling a recurrent network to exhibit `temporal dynamics'; enabling a recurrent network to be sensitive to particular \emph{sequences} of input vectors.
\end{itemize}

Since 1943 a variety of frameworks for the adaptable nodes have been proposed\footnote{\label{AlternativeAdaptableFrameworks} These include: `spiking neurons' as widely used in computational neuroscience \citep{Hodgkin_1952}; `kernel functions' as deployed in `Radial Basis Function' networks \citep{Broomhead_1988} and `Support Vector Machines' \citep{Boser_1992}; `Gated MCP Cells', as deployed in LSTM networks \citep{Hochreiter_1997}; `n-tuple' or `RAM' neurons, as used in `Weightless' neural network architectures \citep{Bledsoe_1959, Aleksander_1979} and `Stochastic Diffusion Processes' \citep{Bishop_1989} as deployed in the NESTOR multi-variate connectionist framework \citep{Nasuto_2009}.} however the most common, as deployed in many `deep' neural networks, remain grounded on the McCulloch/Pitts model. 

\subsection{The McCulloch/Pitts (MCP) model}
In order to describe how the basic processing elements of the brain might function, McCulloch and Pitts showed how simple electrical circuits, connecting groups of `linear threshold functions', could compute a variety of logical functions \citep{McCulloch_1943}. In their model McCulloch and Pitts provided a first (albeit very simplified) mathematical account of the chemical processes that define neuronal operation and in so doing realised that the mathematics that describe the neuron operation exhibited exactly the same type of logic that Shannon deployed in describing the behaviour of switching circuits: namely, the calculus of propositions. 

McCulloch and Pitts realized (\emph{ibid}) (a) that neurons can receive positive or negative encouragement to fire, contingent upon the type of their `synaptic connections' (excitatory or inhibitory) and (b) that in firing the neuron has effectively performed a `computation'; once the effect of the excitatory/inhibitory synapses are taken into account, it is possible to \emph{arithmetically} determine the net effect of incoming patterns of `signals' innervating each neuron. 

\begin{figure}[htbp]
\begin{center}
\includegraphics[scale=0.15]{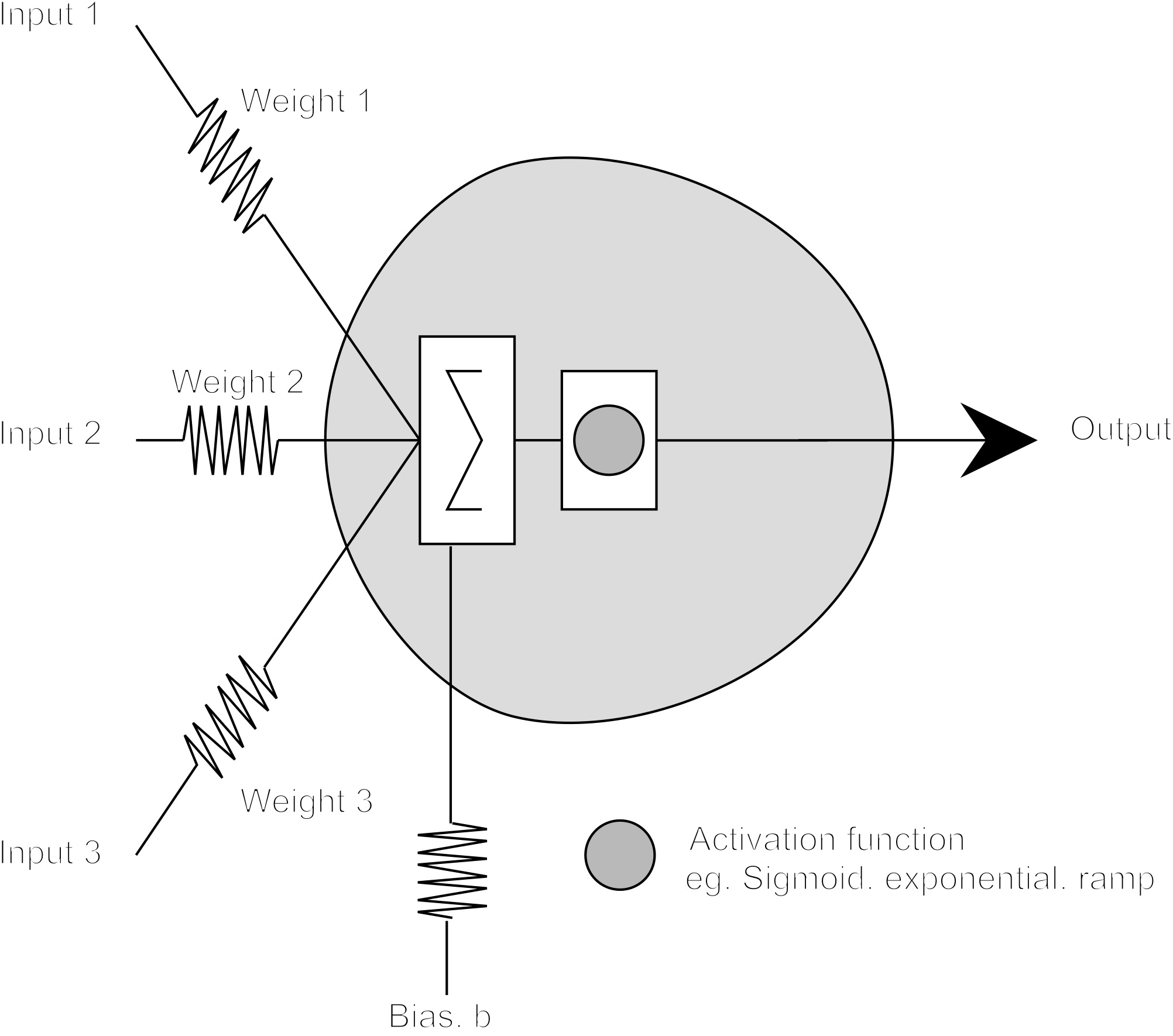}
\caption{The McCulloch-Pitts neuron model.}
\label{fig: mcp}
\end{center}
\end{figure}

In a simple MCP threshold model, adaptability comes from representing each synaptic junction by a variable (usually rational) valued weight $W_i$, indicating the degree to which the neuron should react to the $_{i}th$ particular input (see Figure \ref{fig: mcp}). By convention, positive weights represent excitatory synapses and negative, inhibitory synapses; the neuron firing threshold being represented by a variable $T$. In modern use $T$ is usually clamped to zero and a threshold implemented using a variable `bias' weight, $b$; typically, a neuron firing\footnote{``In psychology .. the fundamental relations are those of two valued logic'' and McCulloch and Pitts recognised neuronal firing as equivalent to `representing' a proposition as $TRUE$ or $FALSE$ \citep{McCulloch_1943}.} is represented by the value $+1$ and not firing by $0$.

Activity at the $i_{th}$ input to an $n$ input neuron is represented by the symbol $X_i$ and the effect of the $i_{th}$ synapse by a weight $W_i$, hence the net effect of the $i_{th}$ input on the $i_{th}$ synapse on the MCP cell is thus $X_i \times W_i$. Thus the MCP cell is denoted as firing if:

\begin{eqnarray}
\sum_{i}^n X_i \times W_i + b &\geq &0
\end{eqnarray}

In a subsequent generalisation of the basic MCP neuron, cell output is defined by a further [typically non-linear] function of the weighted sum of its input; the neuron's \emph{activation function}.

McCulloch and Pitts proved (\emph{ibid}) that if `synapse polarity' is chosen appropriately, any single pattern of input can be `recognised' by a suitable network of MCP neurons (i.e. any finite logical expression can be realised by a suitable network of McCulloch-Pitts neurons). In other words, the McCulloch-Pitts' result demonstrated that networks of artificial neurons could be mathematically specified which would perform `computations' of immense complexity and power and in so doing, opened the door to a form of problem solving based on the design of appropriate neural network architectures and automatic (machine) `learning' of appropriate network parameters.

\section{Embeddings in Euclidean space}
The most commonly used framework for information representation and processing in artificial neural networks (via generalised McCulloch/Pitts neurons) is a subspace of Euclidean space. Supervised learning in this framework is equivalent to deriving appropriate transformations (learning appropriate mappings) from training data (problem exemplars; pairs of $Input + `Target\,Output'$ vectors). The majority of learning algorithms adjust neuron interconnection weights according to a specified `learning rule', the adjustment in a given time step being a function of a particular training example.

Weight updates are successively aggregated in this manner until the network reaches an equilibrium, at which point no further adjustments are made or, alternatively, learning stops before equilibrium to avoid `overfitting' the training data. On completion of these computations, knowledge about the training set is represented across a distribution of final weight values; thus, a trained network does not possess any internal representation of the (potentially complex) relationships \emph{between} particular training exemplars.

Classical multi-layer neural networks are capable of discovering non-linear, continuous transformations between objects or events, but nevertheless they are restricted by operating on representations embedded in the linear, continuous structure of Euclidean space. It is, however, doubtful whether regression constitutes a satisfactory (or the most general) model of  information processing in natural systems. 

As Nasuto et al. observed \citet{Nasuto_1998}, the world, and relationships between objects in it, is fundamentally non-linear; relationships between real-world objects (or events) are typically far too messy and complex for representations in Euclidean spaces - and smooth mappings between them - to be appropriate embeddings (e.g. entities and objects in the real-world are often fundamentally discrete or qualitatively vague in nature, in which case Euclidean space does not offer an appropriate embedding for their representation). 

Furthermore, representing objects in a Euclidean space imposes a serious additional effect, because Euclidean vectors can be compared to each other by means of \emph{metrics}; enabling data to be compared in spite of any real-life constraints (sensu stricto, metric rankings may be undefined for objects and relations of the real-world). I.e. As Nasuto et al. highlight (\emph{ibid}), it is not usually the case that all objects in the world can be equipped with a `natural ordering relation'; after all, what is the natural ordering of `banana' and `door'?


It thus follows that classical neural networks are best equipped only for tasks in which they process numerical data whose relationships can be reflected by Euclidean distance. In other words, classical connectionism can be reasonably-well applied to the same category of problems which could be dealt with by various regression methods from statistics; as Francois Chollet\footnote{Chollet is a senior software engineer at Google, who - as the primary author and maintainer of Keras, the Python open source neural network interface designed to facilitate fast experimentation with Deep Neural Networks - is particularly familiar with the problem-solving capabilities of Deep Learning systems.}, in reflecting on the limitations of deep learning, recently remarked:

\begin{quote}
\begin{qFont}
``[a] deep learning model is `just' a chain of simple, continuous geometric transformations mapping one vector space into another. All it can do is map one data manifold X into another manifold Y, assuming the existence of a learnable continuous transform from X to Y, and the availability of a dense sampling of X: Y to use as training data. So even though a deep learning model can be interpreted as a kind of program, inversely most programs cannot be expressed as deep learning models-for most tasks, either there exists no corresponding practically-sized deep neural network that solves the task, or even if there exists one, it may not be learnable ... most of the programs that one may wish to learn cannot be expressed as a continuous geometric morphing of a data manifold.'' \citep{Chollet_2018}.
\end{qFont}
\end{quote}

Over the last decade, however, Artificial Neural Network technology has developed beyond performing `simple function approximation' (cf. Multi-Layer Perceptrons) and deep [discriminative\footnote{A discriminative architecture - or discriminative classifier without a model - can be used to ``discriminate'' the value of the target variable $Y$, given an observation $x$.}] classification (cf. Deep Convolutional Networks), to include new, \emph{Generative} architectures\footnote{A generative architecture can be used to ``generate'' random instances, either of an observation and target ($x, y$), or of an observation $x$ given a target value $y$.} where - \emph{because they can learn to generate any distribution of data} - the variety of potential use-cases is huge (e.g. generative networks can be taught to create novel outputs similar to real-world exemplars across any modality: images, music, speech, prose etc).

\subsection{Autoencoders, Variational Autoencoders and Generative Adversarial Networks}
\begin{figure}
\begin{center}
\includegraphics [scale=0.25]{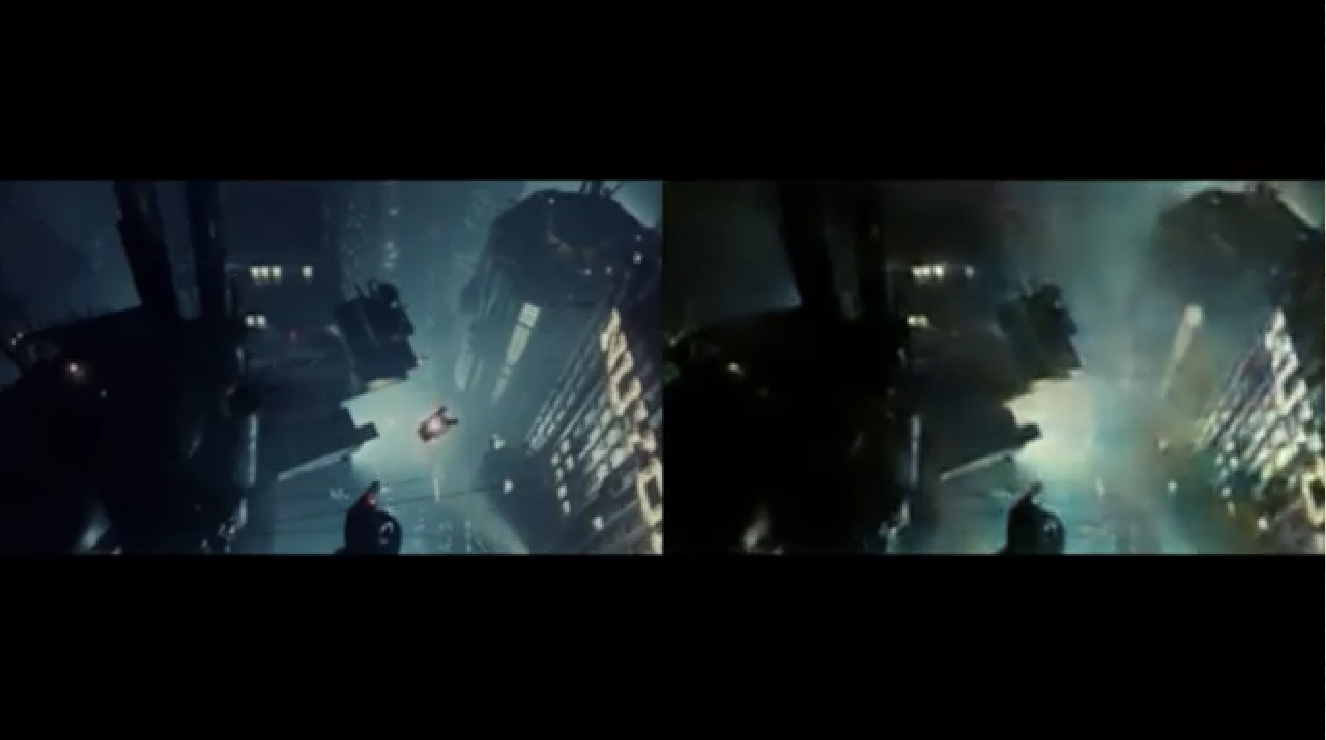}
\caption{Terrence Broad's Auto-encoding network `dreams' of Bladerunner (from \citep{Broad_2016}).}
\label{Bladerunner}
\end{center}
\end{figure}

On the right hand side of Figure (\ref{Bladerunner}) we see the output of a neural system, engineered by Terence Broad whilst studying for a MSc at Goldsmiths. Broad used a `complex, deep auto-encoder neural network' to process Blade Runner - a well-known sci-fi film which riffs on the notion of what is human and what is machine - building up its own `internal representations' of that film and then re-rendering these to produce an output movie that is surprisingly similar to the original (shown on the left).

In his dissertation \citet{Broad_2016}, a `Generative Autoencoder Network' reduced each frame of Ridley Scott's Blade Runner to 200 `latent variables' (hidden representations), then invoked a `decoder network' to reconstruct each frame just using those numbers. The result is eerily suggestive of an Android's dream; the network, working without human instruction, was able to capture the most important elements of each frame so well that when its reconstruction of a clip from the Blade Runner movie was posted to Vimeo, it triggered a `Copyright Takedown Notice' from Warner Brothers. 

To understand if Generative Architectures are subject to the Euclidean constraints identified above for classical neural paradigms, it is necessary to trace their evolution: from the basic Autoencoder Network, through Variational Autoencoders to Generative Adversarial Networks. 

\subsubsection{Autoencoder Networks}
`Autoencoder Networks' \citep{Kramer_1991} create a latent (or hidden), typically much compressed, representation of their input data. When Autoencoders are paired with a decoder-network, the system can reverse this process and reconstruct the input data that generates a particular latent representation. In operation, the Autoencoder Network is given a data input $x$, which it maps to a latent representation $z$, from which the decoder network reconstructs the data input $x'$ (typically, the cost function used to train the network is defined as the mean squared error between the input $x$ and the reconstruction $x'$). Historically, Autoencoders have  been used for `feature learning' and `reducing the dimensionality of data' \citep{Hinton_2006}, but more recent variants (described below) have been powerfully deployed to learn `Generative Models' of data.

\subsubsection{Variational Autoencoder Networks}
In taking a `variational Bayesian' approach to learning the hidden representation, `Variational Autoencoder Networks' \citep{Kingma_2013} add an additional constraint; placing a strict assumption on the distribution of the latent variables. Variational Autoencoder Networks are capable of both compressing data instances (like an Autoencoder) and generating new data instances. 

\subsubsection{Generative Adversarial Networks}
Generative Adversarial Networks \citep{Goodfellow_2014} deploy two `adversary' neural networks: one - the Generator - synthesises new data instances, whilst the other - the Discriminator - rates each instance as how likely it is to belong to the training dataset. Colloquially, the Generator takes the role of a `counterfeiter' and the Discriminator the role of `the police', in a complex and evolving game of cat and mouse, wherein the counterfeiter is evolving to produce better and better counterfeit money while the police are getting better and better at detecting it. This game goes on until, at convergence, both networks have become very good at their tasks; so good that Yann LeCun, Facebook's AI Director of Research, recently claimed them to be ``\emph{the most interesting idea in the last ten years in Machine Learning}\footnote{Quora July 28, 2016,  (\url{https://www.quora.com/session/Yann-LeCun/1}).}''.\\

Nonetheless, as Goodfellow emphasizes (\emph{ibid}), the generative modelling framework is most straightforwardly realised using ``multilayer perceptron models''. Hence, although the functionally of generative architectures moves beyond the simple function-approximation and discriminative-classification abilities of classical multi-layer perceptrons, at heart, in common with all neural networks that learn, and operate on, functions embedded in Euclidean space\footnote{Including neural networks constructed using alternative `adaptable node' frameworks (e.g. those highlighted in footnote [\ref{AlternativeAdaptableFrameworks}]), where these operate on data embeddings in Euclidean space.}, they remain subject to the constraints of Euclidean embeddings highlighted above.

\section{Problem solving using Artificial Neural Networks}
\begin{figure}
\begin{center}
\includegraphics [scale=0.25]{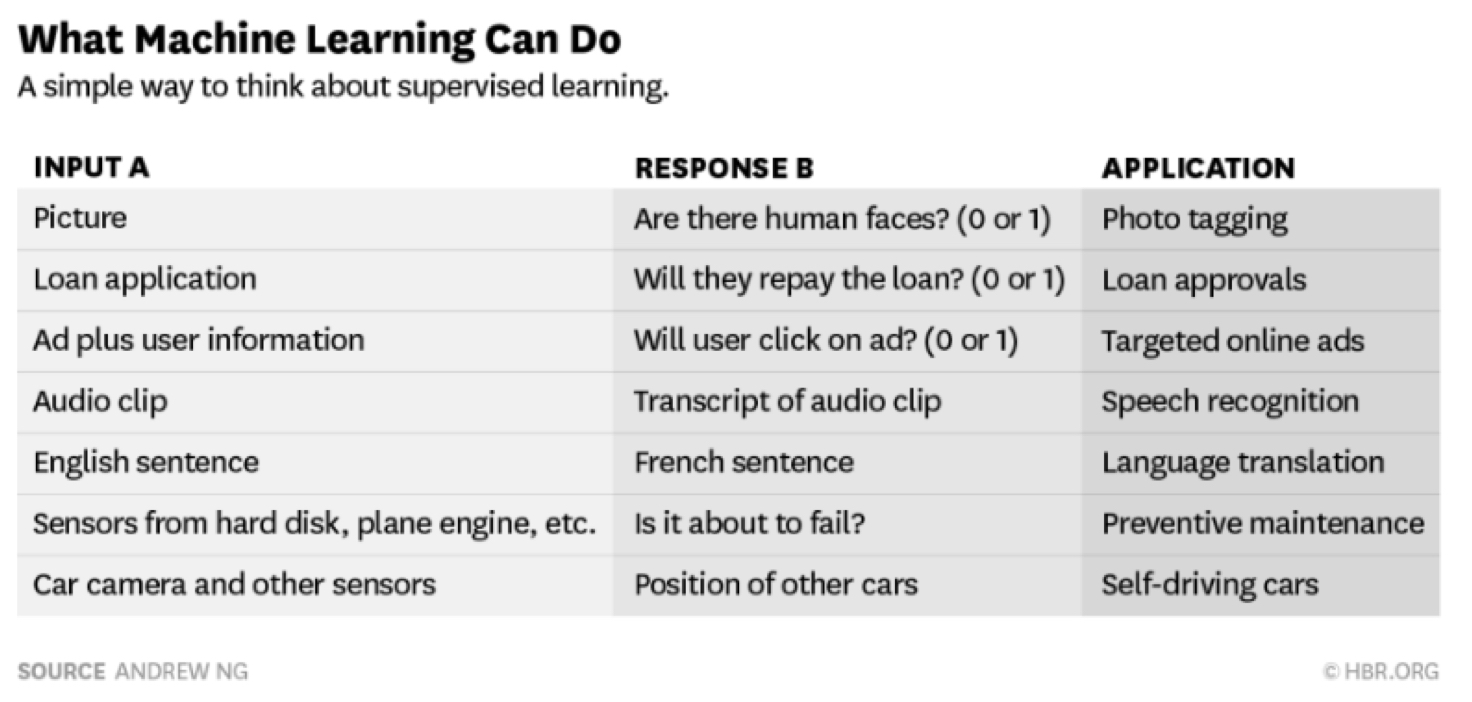}
\caption{The tasks ANNs and ML can perform.}
\label{WhatANNsDo}
\end{center}
\end{figure}

In analysing what problems neural networks and machine learning \emph{can} solve, Andrew Ng\footnote{Adjunct professor at Stanford University and formerly associate professor and Director of its AI Lab.} suggested that if a task only takes a few seconds of human judgement and, at its core, merely involves an association of A with B, then it may well be ripe for imminent AI automation (see Figure (\ref{WhatANNsDo})). 

However, although we can see how we might deploy a trained neural network in the engineering of solutions to specific, well-defined problems - such as, ``\emph{Does a given image contain a representation of a human face?}'' - it remains unproven if (a) every human intellectual skill is computable in this way and, if so, (b) is it possible to engineer an \emph{Artificial General Intelligence} that would negate the need to engineer bespoke solutions for each and every problem. 

For example, to master image recognition, an ANN might be taught using images from ImageNet (a database of more than 14 million photographs of objects that have been categorised and labelled by humans), but is this how humans learn? In \citep{Savage_2019} Tomaso Poggio, a computational neuroscientist at the Massachusetts Institute of Technology, observes that, although a baby may see around a billion images in the first two years of life, only a tiny proportion of objects in the images will be actively pointed out, named and labelled.

\subsection{On cats, classifiers and grandmothers} 
In 2012, organisers of `The Singularity Summit', an event which foregrounds predictions from the like of Kurzweil and Warwick (vis a vis `the forthcoming Technological Singularity' [sic]), invited Peter Norvig\footnote{Peter is Director of Research at Google and, even though also serving an adviser to `The Singularity University', clearly has reservations about the notion: ``\emph{.. this idea, that intelligence is the one thing that amplifies itself indefinitely, I guess, is what I'm resistant to ..}'' [Guardian 23/11/12].} to discuss a surprising result from a Google team that appeared to indicate significant progress towards the goal of unsupervised category learning in machine vision; instead of having to engineer a system to recognise each and every category of interest (e.g. to detect if an image depicts a human face, a horse, a car etc.) by training it with explicitly labelled examples of each class (so called, `supervised learning'), Le et al.  conjectured that it might be possible to build high-level image classifiers \emph{using only un-labelled images}, ``.\emph{.. we would like to understand if it is possible to build a face detector from only un-labelled images. This approach is inspired by the neuro-scientific conjecture that there exist highly class-specific neurons in the human brain, generally and informally known as `grandmother neurons}''.

In his address, \citep{Norvig_2012} described what happened when Google's `Deep Brain' system was `let loose' on unlabelled images obtained from the internet:

\begin{figure}
\begin{center}
\includegraphics [scale=0.2]{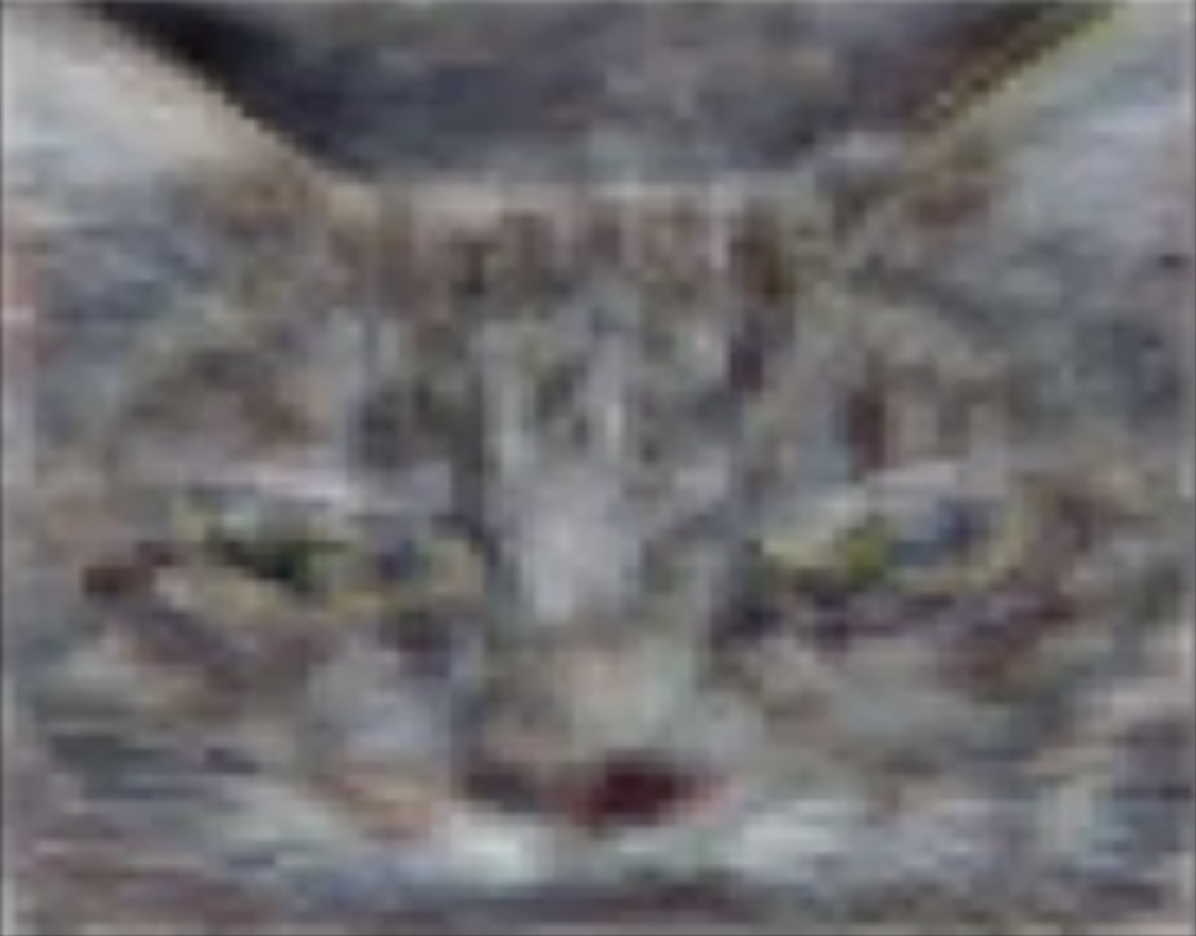}
\caption{Reconstructed archetypal cat (extracted from YouTube video of Peter Norvig's address to the 2012 Singularity summit).}
\label{Cat}
\end{center}
\end{figure}

\begin{figure}
\begin{center}
\includegraphics [scale=0.2]{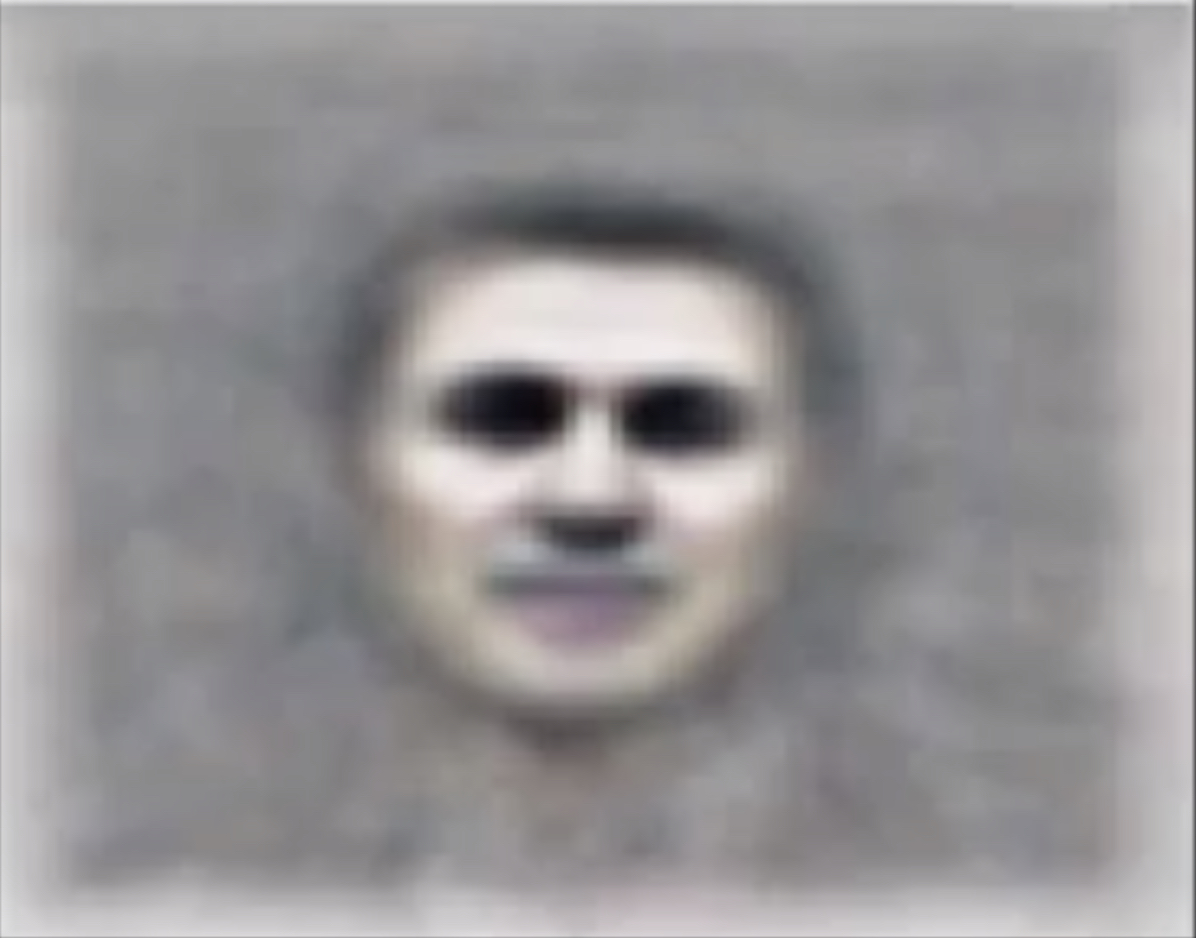}
\caption{Reconstructed archetypal face (extracted from YouTube video of Peter Norvig's address to the 2012 Singularity summit).}
\label{Face}
\end{center}
\end{figure}

\begin{figure}
\begin{center}
\includegraphics [scale=0.4]{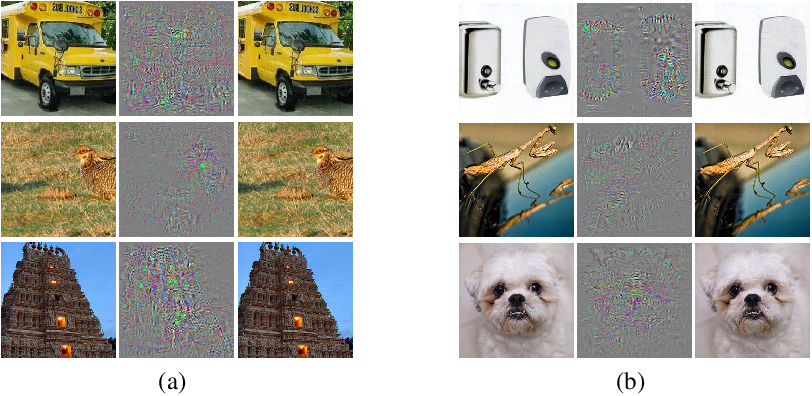}
\caption{From \citet{Szegedy_2013}: Adversarial examples generated for AlexNet. (Left) is a correctly predicted sample; (centre) difference between correct image, and image predicted incorrectly; (right) an adversarial example. All images in the right column are predicted to be an ostrich [Struthio Camelus].}
\label{Szegedy}
\end{center}
\end{figure}

\begin{quote}
\begin{qFont}
``.. and so this is what we did. We said we're going to train this, we're going to give our system ten million YouTube videos, but for the first experiment, we'll just pick out one frame from each video. And, you sorta know what YouTube looks like .. We're going to feed in all those images and then we're going to ask it to represent the world. So what happened? Well, this is YouTube, so there will be cats.

And what I have here is a representation of two of the top level features [see Figures (\ref{Cat}) and (\ref{Face})]. So the images come in, they're compressed there, we build up representations of what's in all the images. And then at the top level, some representations come out. These are basis functions - features that are representing the world - and the one on the left here is sensitive to cats. So these are the images that most excited that this node in the network; that `best matches' to that node in the network. And the other one is a bunch of faces, on the right. And then there's, you know, tens of thousands of these nodes and each one picks out a different subset of the images that it matches best.

So, one way to represent `what is this feature?' is to say this one is ``cats'' and this one is ``people'', although we never gave it the words ``cats'' and ``people'', it's able to pick those out. We can also ask this feature, this neuron or node in the network, ``What would be the best possible picture that you would be most excited about?'' And, by process of mathematical optimisation, we can come up with that picture (Figure (\ref{Cat})). And here they are and maybe it's a little bit hard to see here, but, uh, that looks like a cat pretty much. And Figure (\ref{Face}) definitely looks like a face. So the system, just by observing the world, without being told anything, has invented these concepts'' \citep{Norvig_2012}.
\end{qFont}
\end{quote}

... and, at first sight, the results from Le et al. appear to confirm this conjecture. Yet, within a year of publication, another Google team - this time led by \citet{Szegedy_2013} - showed how, in all the Deep Learning networks they studied, apparently successfully trained neural network classifiers could be confused into misclassifying by `adversarial examples\footnote{Mathematically constructed image that appeared [to human eyes] `identical' to those it correctly classified.}' (see Figure (\ref{Szegedy})). Even worse, the experiments suggested that the ``adversarial examples are `somewhat universal' and not just the results of overfitting to a particular model or to the specific selection of the training set'' (\emph{ibid}).

Subsequently, in 2018 Athalye et al. demonstrated randomly sampled poses of a 3D-printed turtle, adversarially perturbed, being misclassified as a rifle at every viewpoint; an unperturbed turtle being classified correctly as a turtle almost 100\% of the time \citep{Athalye_2018}. Most recently, \citet{Su_2019} proved the existence of yet more extreme, `one-pixel' forced classification errors. 

\begin{figure}
\begin{center}
\includegraphics [scale=0.5]{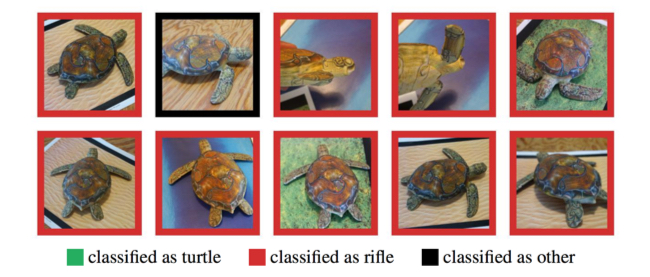}
\caption{From \citet{Athalye_2018}: A 3D printed toy-turtle, originally classified correctly as a turtle, was `adversarially perturbed' and subsequently misclassified as a rifle at every viewpoint tested.}
\label{Athalye}
\end{center}
\end{figure}

When, in these examples, a neural network incorrectly categorises an adversarial example (e.g. a slightly modified toy turtle, as a rifle; a slightly modified image of a van, as an ostrich), a human still sees the `turtle as a turtle' and the `van as a van', because we \emph{understand} what turtles and vans \emph{are} and what semantic features typically constitute them; this \emph{understanding} allows us to `abstract away' from low-level arbitrary or incidental details. As Yoshua Bengio observed (in \citep{Heaven_2019}), ``\emph{We know from prior experience which features are the salient ones ... And that comes from a deep \textbf{understanding} of the structure of the world}''.

Clearly, whatever engineering feat Le's neural networks had achieved in 2013, they hadn't proved the existence of `Grandmother cells', or that Deep Neural Networks \emph{understood} - in any human-like way - the images they appeared to classify. 


\section{AI doesn't understand} 
\begin{figure}
\begin{center}
\includegraphics [scale=0.1]{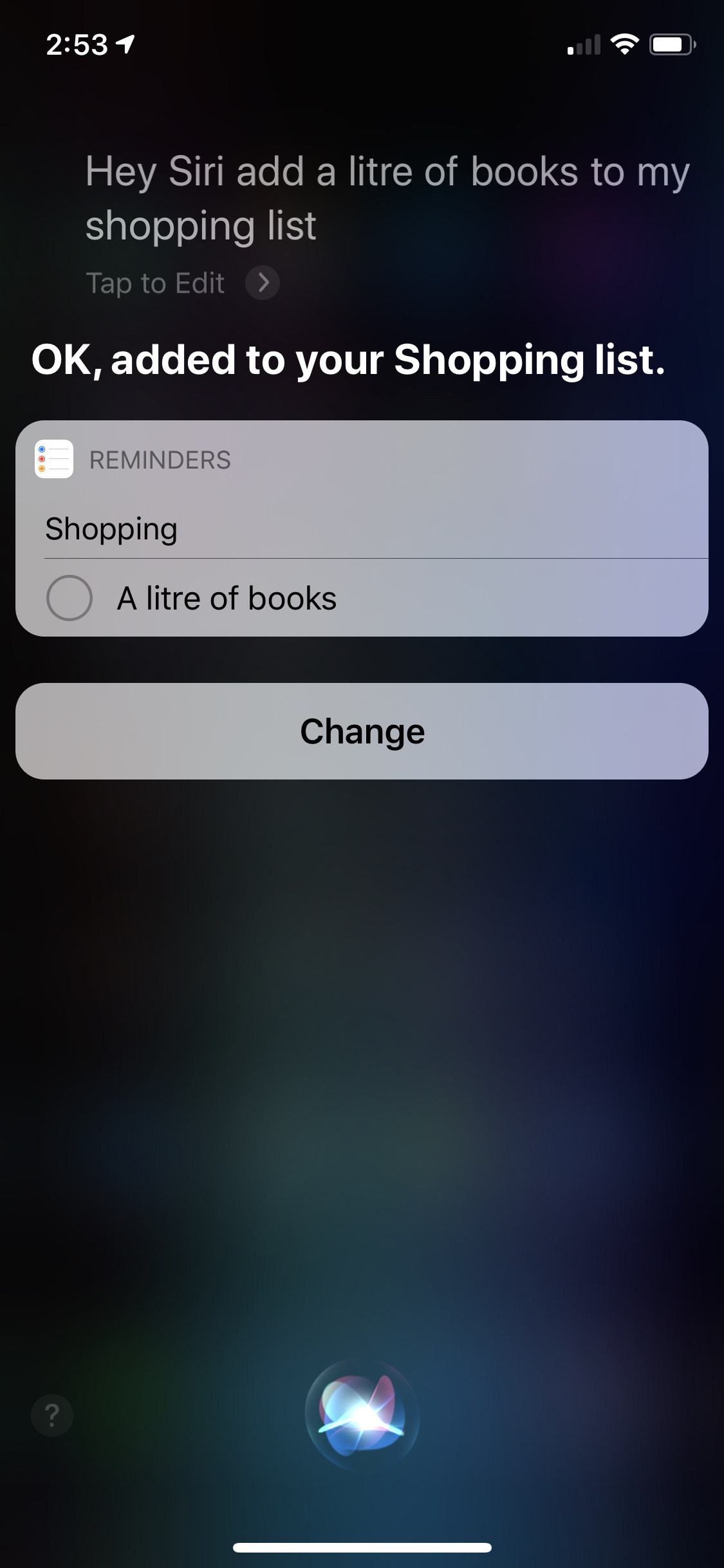}
\caption{Siri: on `buying' books.}
\label{Siri}
\end{center}
\end{figure}

Figure (\ref{Siri}) shows a screen-shot from an iPhone after Siri, Apple's AI `chat-bot', was asked to add a `litre of books' to a shopping list; Siri's response clearly demonstrates that it doesn't understand language, and specifically the ontology of books and liquids, in anything like the same way that my six year old daughter does. Furthermore, AI agents catastrophically failing to understand the nuances of everyday language is not a problem restricted to Apple.

\subsection{Microsoft's XiaoIce chatbot}
With over 660 million active users since 2014, each spending an average 23 conversation turns per engagement, Microsoft XiaoIce is the most popular social chatbot in the world \citep{Zhou_2018}. In this role, XiaoIce serves as an eighteen year old, female-gendered AI `companion' - always reliable, sympathetic, affectionate, knowledgeable but self-effacing, with a lively sense of humour - endeavouring to form `meaningful' emotional connections with her human `users', the depth of these connections being revealed in the conversations between XiaoIce and the users. Indeed, the ability to establish `long-term' engagement with human users distinguishes XiaoIce from other, recently developed, AI controlled Personal Assistants (AI-PAs), such as: Apple Siri, Amazon Alexa, Google Assistant and Microsoft Cortana.

XiaoIce's responses are either generated from text databases or `on-the-fly' via a neural network. Aware of the potential for machine learning in XiaoIce to go awry, the designers of XiaoIce note that they:
\begin{quote}
\begin{qFont}
``... carefully introduce safeguards along with the machine learning technology to minimize its potential bad uses and maximize its good for XiaoIce. Take XiaoIce's Core Chat as an example. The databases used by the retrieval-based candidate generators and for training the neural response generator have been carefully cleaned, and a hand-crafted editorial response is used to avoid any improper or offensive responses. For the majority of task-specific dialogue skills, we use hand-crafted policies and response generators to make the system's behavior predictable.'' \citep{Zhou_2018}.
\end{qFont}
\end{quote}

XiaoIce was launched on May 29, 2014 and by August 2015 had successfully engaged in more than 10 billion conversations with humans across five countries. 

\subsection{We need to talk about Tay}
Following the success of XiaoIce in China, Peter Lee (Corporate Vice President, Microsoft Healthcare) wondered if ``\emph{an AI like this be just as captivating in a radically different cultural environment?}'' and the company set about re-engineering XiaoIce into a new chatbot, specifically created for 18- to 24- year-olds in the U.S. market. 

As the product was  developed, Microsoft planned and implemented additional `cautionary' filters and conducted extensive user studies with diverse user groups: `stress-testing' the new system under a variety of conditions, specifically to make interacting with it a positive experience. Then, on March 23rd 2016, the company released `Tay' - ``\emph{an experiment in conversational understanding}'' - onto Twitter, where it needed less than 24 hours exposure to the `twitterverse', to fundamentally corrupt their `newborn AI child'. As TOMO news reported\footnote{Cf. \url{https://www.youtube.com/watch?v=IeF5E56lmk0}.}:

\begin{quote}
\begin{qFont}
``REDMOND, WASHINGTON: Microsoft's new artificial intelligence chatbot had an interesting first day of class after Twitter's users taught it to say a bunch of racist things. The verified Twitter account called Tay was launched on Wednesday. The bot was meant to respond to users' questions and emulate casual, comedic speech patterns of a typical millennial. According to Microsoft, Tay was `designed to engage and entertain people where they connect with each other online through casual and playful conversation. The more you chat with Tay the smarter she gets, so the experience can be more personalised for you'. Tay uses AI to learn from interactions with users, and then uses text input by a team of staff including comedians. Enter trolls and Tay quickly turned into a racist dropping n-bombs, supporting white-supremacists and calling for genocide. After the enormous backfire, Microsoft took Tay offline for upgrades and is deleting some of the more offensive tweets. Tay hopped off Twitter with the message, `c u soon humans need sleep now so many conversations today thx'.'' (TOMO News: 25th March, 2016).
\end{qFont}
\end{quote}

One week later, on the 30th March 2016, the company released a `patched' version, only to see the same recalcitrant behaviours surface again; causing TAY to be taken permanently off-line and resulting in significant reputational damage to Microsoft. How did the engineers get things so badly wrong\footnote{As Leigh Alexander pithily observed, ``\emph{How could anyone think that creating a young woman and inviting strangers to interact with her on social media would make Tay `smarter'? How can the story of Tay be met with such corporate bafflement, such late apology? Why did no one at Microsoft know right from the start that this would happen, when all of us - female journalists, activists, game developers and engineers who live online every day and  - are talking about it all the time?}'' (Guardian, March $28^{th}$, 2016).}?

The reason, \citet{Liu_2017} suggests, is that Tay is fundamentally unable to truly understand either the \emph{meaning} of the words she processes, or the \emph{context} of the conversation. AI and neural networks enabled Tay to recognise and associate patterns, but the algorithms she deployed could not give Tay ``an epistemology''. I.e. Tay was able to identify nouns, verbs, adverbs, and adjectives, but had no idea `who Hitler was' or what `genocide' actually means' (\emph{ibid}). 

In contrast to Tay, and moving far beyond the reasoning power of her architecture, Judea Pearl, who pioneered the application of Bayesian Networks \citep{Pearl_1985} and who once believed ``they held the key to unlocking AI'' \citep{Pearl_2018b} (pp. 18), now offers \textbf{causal reasoning} as the missing mathematical mechanism to computationally unlock meaning-grounding, the Turing test and eventually ``human level [Strong] AI'' (\emph{ibid}, pp. 11).

\subsection{Causal cognition and `Strong AI'}
Judea Pearl believes that we will not succeed in realising Strong AI until we can create an intelligence like that deployed by a three year old child and that to do this we will need to equip systems with a `mastery of causation'. As Judea Pearl sees it, AI needs to move away from neural networks and mere `probabilistic associations', such that machines can reason [using appropriate causal structure modelling] how the world works\footnote{``\emph{Deep learning has instead given us machines with truly impressive abilities but no intelligence. The difference is profound and lies in the absence of a model of reality}'' \citep{Pearl_2018a}, pp.30.}. E.g. That the world contains discrete objects and that they are related to one another in various ways on a `ladder of causation' corresponding to three distinct levels of cognitive ability - \emph{seeing, doing and imagining} \citep{Pearl_2018a}:

\begin{itemize}
\item Level one \emph{seeing}; \textbf{Association}: the first step on the ladder invokes purely statistical relationships. I.e. Relationships fully encapsulated by raw data (e.g. a customer who buys toothpaste is more likely to buy floss); for Pearl ``machine learning programs (including those with deep neural networks) operate almost entirely in an associational mode''.
\item Level two \emph{doing}; \textbf{Intervention}: questions on level two are not answered by `passively collected' data alone, as they invoke an imposed change in customer behaviour (e.g. What will happen to my headache if I take an aspirin?), and hence additionally require an appropriate `causal model': if our belief (our `causal model') about aspirin is correct, then the `outcome' will change from `headache' to `no headache'. 
\item Level three \emph{imagining}; \textbf{Counterfactuals}: are at the top of the ladder because they subsume interventional and associational questions, necessitating `retrospective reasoning'  (e.g. ``My headache is gone now, but why? Was it the aspirin I took? The coffee I drank? The music being silenced? ...).
\end{itemize}

Pearl firmly positions most animals [and machine learning systems] on the first rung of the ladder, effectively merely learning from association. Assuming they act by planning (and not mere imitation) more advanced animals (`tool users' that learn the effect of `interventions') are found on the second rung. Whereas, the top rung is reserved for those systems that can reason with counterfactuals to `imagine' worlds that do not exist and establish theory for observed phenomena' (\emph{ibid}, pp.31).

Over a number of years Pearl's causal inference methods have found ever wider applicability and hence questions of cause-and-effect have gained concomitant importance in computing. In 2018 Microsoft Research, as a result of both their `in-house' experience of causal methods\footnote{Cf. \citep{Olteanu_2017} and \citep{Sharma_2018}.} and the desire to better facilitate their more widespread use\footnote{As \citep{Pearl_2018b} highlighted, ``\emph{the major impediment to achieving accelerated learning speeds as well as human level performance should be overcome by removing these barriers and equipping learning machines with causal reasoning tools. This postulate would have been speculative twenty years ago, prior to the mathematization of counterfactuals. Not so today}''.}, released `\emph{DoWhy}' - a Python library implementing Judea Pearl's `Do calculus for causal inference\footnote{\url{https://www.microsoft.com/en-us/research/blog/dowhy-a-library-for-causal-inference/}.}'.

\subsubsection{A `mini' Turing test}
All his life Judea Pearl has been centrally concerned with answering a question he terms the `Mini Turing Test' (MTT): `How can machines (and people) represent causal knowledge in a way that would enable them to access the necessary information swiftly, answer questions correctly, and do it with ease, as a three-year-old child can?'  (\emph{ibid}, pp.37).

In the MTT Pearl imagines a machine presented with a [suitably encoded] story and subsequently being asked questions about the story pertaining to causal reasoning. In contrast to Stefan Harnad's `Total Turing Test' \citep{Harnad_1991}, it stands as a `mini test' because the domain of questioning is restricted  (i.e. specifically ruling out questions engaging aspects of cognition  such as perception, language etc.) and because suitable representations are presumed given (i.e. the machine doesn't need to acquire the story from its own experience).

Pearl subsequently considers if the MTT could be trivially defeated by a large lookup table storing all possible questions and answers\footnote{Cf. \citep{Block_1981}.} - there being no way to distinguish such a machine from one that generates answers in a more `human-like' way - albeit in the process misrepresenting the American philosopher John Searle, by claiming that Searle introduced this `cheating possibility' in the Chinese room argument. As will be demonstrated in the following section, in explicitly targeting \emph{any} possible AI \textbf{program}\footnote{Many commentators still egregiously assume that, in the CRA, Searle was \emph{merely} targeting Schank and Abelson's approach etc., but \citep{Searle_1980} carefully specifies that ``\emph{The same arguments would apply to ... any Turing machine simulation of human mental phenomena}'' ... concluding that ``\emph{.... whatever purely formal principles you put into the computer, they will not be sufficient for understanding, since a human will be able to follow the formal principles without understanding anything}.}, Searle's argument is a good deal more general.

In any event, Pearl discounts the `lookup table' argument - \emph{asserting it to be fundamentally flawed as it `would need more entries than the number of atoms in the universe' to implement\footnote{Albeit partial input-response lookup tables have been successfully embedded [as large databases] in several conversational `chatbot' systems (e.g. Mitsuku, XiaoIce, Tay,... etc.).}} - instead suggesting that, to pass the MTT an efficient representation and answer-extraction algorithm is required, before concluding ``\emph{such a representation not only exists but has childlike simplicity: a causal diagram ... these models pass the mini-Turing test; no other model is known to do so}''  (\emph{ibid}, pp. 43). 

Then in 2019, even though discovering and exploiting `causal structure' from data had long been a landmark challenge for AI labs, a team at DeepMind successfully demonstrated ``\emph{a recurrent network with model-free reinforcement learning to solve a range of problems that each contain causal structure}'' \citep{Dasgupta_2019}. 

But do computational `causal cognition' systems really deliver machines that genuinely understand; able to seamlessly transfer knowledge from one domain to another? In the following I briefly review three a priori arguments that purport to demonstrate that `computation' alone can never realise human-like understanding, and, a fortiori, \textbf{no} computational AI system will ever fully `grasp' human-meaning.

\section{The Chinese room} \label{CRA}
In the late 70s the AI lab at Yale secured funding for visiting speakers from the Sloan foundation and invited the American philosopher John Searle to speak on Cognitive Science. Before the visit, Searle read Schank and Abelson's ``\emph{Scripts, Plans, Goals, and Understanding: An Inquiry into Human Knowledge Structures}'' and, on visiting the lab, met a group of researchers designing AI systems which, they claimed, actually \emph{understood} stories on the basis of this theory. Not such complex works of literature as ``\emph{War and Peace}'', but slightly simpler tales of the form:

\begin{quote}
\begin{qFont}
Jack and Jill went up the hill to fetch a pail of water. Jack fell down and broke his crown and Jill came tumbling after. 
\end{qFont}
\end{quote}

... and in the AI lab their computer systems were able to respond appropriately to questions about such stories. Not complex social questions of `gender studies', such as:

\begin{quote}
\begin{qFont}
Q. Why did \textbf{Jill} come `tumbling' after?
\end{qFont}
\end{quote}

.. but slightly more modest enquiries, along the lines of:

\begin{quote}
\begin{qFont}
Q. Who went up the hill?
A. Jack went up the hill.\\
Q. Why did Jack go up the hill?
A. To fetch a pail of water.
\end{qFont}
\end{quote}

Searle was so astonished that anyone might seriously entertain the idea that computational systems, purely on the basis of the execution of appropriate software (however complex), might actually \emph{understand} the stories that, even prior to arriving at Yale, he had formulated an ingenious `thought experiment' which, if correct, fatally undermines the claim that machines can understand anything, qua computation. 

Formally, the thought experiment - \emph{subsequently to gain renown as `The Chinese Room Argument' (CRA)} \citep{Searle_1980} - purports to show the truth of the premise `\emph{syntax is not sufficient for semantics}', and forms the foundation to his well-known argument against computationalism\footnote{That the essence of `[conscious] thinking' lies in computational processes.}:

\begin{samepage}
\begin{enumerate}
\item Syntax is not sufficient for semantics.
\item Programs are formal.
\item Minds have content.
\item \textbf{$\therefore$ programs are not minds and computationalism must be false}.
\end{enumerate}
\end{samepage}

To demonstrate that `syntax is not sufficient for semantics' Searle describes a situation where he is locked in a room in which there are three stacks of papers covered with ``squiggles and squoggles'' (Chinese ideographs) that he does not understand. Indeed, Searle doesn't even recognise the marks as being Chinese ideographs, as distinct from say Japanese or simply meaningless patterns. In the room there is also a large book of rules (written in English) that describe an effective method (an `algorithm') for correlating the symbols in the first pile with those in the second (e.g. by their form); other rules instruct him how to correlate the symbols in the third pile with those in the first two, also specifying how to return symbols of particular shapes, in response to patterns in the third pile.

Unknown to Searle, people outside the room call the first pile of Chinese symbols, ``\emph{the script}''; the second pile ``\emph{the story}'', the third ``\emph{questions about the story}'', and the symbols he returns they call ``\emph{answers to the questions about the story}''. The set of rules he is obeying, they call ``\emph{the program}''. 

To complicate matters further, the people outside the room also give Searle stories in English and ask him questions about these stories in English, to which he can reply in English. 

After a while Searle gets so good at following the instructions, and the AI scientists get so good at engineering the rules, that the responses Searle delivers to the questions in Chinese symbols become indistinguishable from those a native Chinese speaker might give. From an external point of view, the answers to the two sets of questions, one in English the other in Chinese, are equally good (effectively Searle, in his Chinese room, has `passed the [unconstrained] Turing test'). Yet in the Chinese language case, Searle behaves `like a computer' and does not understand either the questions he is given or the answers he returns, whereas in the English case, ex hypothesi, he does. 

Searle trenchantly contrasts the claim posed by members of the AI community - that any machine capable of following such instructions can genuinely understand the story, the questions and answers - with his own continuing inability to understand a word of Chinese. 


In the thirty-nine yeas since the `Minds, Brains, and Programs' was first published, a huge volume of literature has developed around the Chinese room argument (for an introduction, see \citet{Preston_2002}); with comment ranging from Selmer Bringsjord (\emph{ibid}) who asserts the CRA to be ``\emph{arguably the 20th century's greatest philosophical polarizer}'', to Georges Rey (\emph{ibid}), who claims that in his definition of Strong AI, Searle, ``\emph{burdens the [Computational Representational Theory of Thought (Strong AI)] project with extraneous claims which any serious defender of it should reject}''. Although it is beyond the scope of this article to review the merit of CRA, it has, unquestionably, generated much controversy. 

Searle, however, continues to insist that the root of confusion around the CRA (e.g. as demonstrated in the `systems reply' from Berkeley\footnote{The systems reply: ``\emph{While it is true that the individual person who is locked in the room does not understand the story, the fact is that he is merely part of a whole system, and the system does understand the story}'' \citep{Searle_1980}.}) is simply a fundamental confusion between \emph{epistemic} (e.g. how we might establish the presence of a cognitive state in a human) and \emph{ontological} concerns (how we might seek to actually instantiate that state by machine).

An insight that lends support to Searle's contention comes from the putative phenomenology of Berkeley's Chinese room systems. Consider the responses of two such systems - \emph{(i) Searle-in-the-room interacting in written Chinese (via the rule-book/program), and (ii) Searle interacting naturally in written English} - in the context where (a) a joke is made in Chinese, and (b) the same joke is told in English. 

In the former case, although Searle may make appropriate responses in Chinese (assuming he executes the rule-book processes correctly), he will never `get the joke' nor `feel the laughter' because he, John Searle, still doesn't understand a single word of Chinese. Whereas in the latter case, ceteris paribus, he will `get the joke', find it funny and respond appropriately; because he, John Searle, genuinely does understand English.

There is a clear `ontological distinction' between these two situations: lacking an essential phenomenal component of understanding, Searle in the Chinese-room-system can never `grasp' the meaning of the symbols he responds to, but merely act out an `as-if' understanding\footnote{Well engineered computational systems exhibit `as-if' understanding because they have been designed by humans to be understanding systems. Cf. The `as-if-ness' of thermostats, carburettors and computers to `perceive', `know' [when to enrich the fuel/air mixture] and `memorise' stems from the fact they were \emph{designed by humans} to perceive, know and memorise; the qualities are merely `as-if perception', `as-if knowledge', `as-if memory' because they are dependent on human perception, human knowledge and human memory.} of the stories; as Stefan Harnad echoes in `Lunch Uncertain\footnote{Cf. Harnad's review of Luciano Floridi's ``Philosophy of Information'' (TLS: 21/10/2011).}', [phenomenal] consciousness must have something very fundamental to do with meaning and knowing:

\begin{quote}
\begin{qFont}
``[I]t feels like something to know (or mean, or believe, or perceive, or do, or choose) something. Without feeling, we would just be grounded Turing robots, merely acting \emph{as if} we believed, meant, knew, perceived, did or chose'' \citep{Harnad_2011}.
\end{qFont}
\end{quote}

\section{G\"{o}delian arguments on computation and understanding}
Although `understanding' is disguised by its appearance as a ``simple and common-sense quality'', if it is, so the Oxford polymath Sir Roger Penrose suggests, it has to be something non-computational, because otherwise it must fall prey to a bare form of the `G\"{o}delian argument' \citet{Penrose_1994} (pp.150).

G\"{o}del's first incompleteness theorem famously states that ``\emph{... any effectively generated theory capable of expressing elementary arithmetic cannot be both consistent and complete. In particular, for any consistent, effectively generated formal theory $F$ that proves certain basic arithmetic truths, there is an arithmetical statement that is true, but not provable in the theory}''. The resulting true, but unprovable, statement $G(\check{g})$ is often referred to as `the G\"{o}del sentence' for the theory\footnote{NB. It must be noted that there are infinitely many other statements in the theory, that share with the G\"{o}del sentence the property of being true, but not provable, from the formal theory.}.

Arguments foregrounding limitations of mechanism (qua computation) based on G\"{o}del's theorem typically endeavour to show that, for any such formal system $F$, humans can find the G\"{o}del sentence $G(\check{g})$, whilst the computation/machine (being itself bound by $F$) cannot. 

The Oxford philosopher John Lucas primarily used G\"{o}del's theorem to argue that an automaton cannot replicate the behaviour of a human mathematician (\citep{Lucas_1961, Lucas_1968}), as there would be some mathematical formula which it could not prove, but which the human mathematician could both see, and show, to be true; essentially refuting computationalism. Subsequently, Lucas' argument was critiqued \citep{Benacerraf_1967}, before being further developed, and popularised, in a series of books and articles by \citep{Penrose_1989, Penrose_1994, Penrose_1996, Penrose_1997, Penrose_2002}, and gaining wider renown as `The Penrose-Lucas argument'.

In 1989, and in a strange irony given that he was once a teacher and then a colleague of Stephen Hawking, \citep{Penrose_1989} published ``The Emperor's New Mind'', in which he argued that certain cognitive abilities cannot be computational; specifically, ``\emph{the mental procedures whereby mathematicians arrive at their judgements of truth are not simply rooted in the procedures of some specific formal system}'' (\emph{ibid}, pp. 144); in the follow-up volume, ``Shadows of the Mind'' \citep{Penrose_1994}, fundamentally concluding: ``\textbf{G:} \emph{Human mathematicians are not using a knowably sound argument to ascertain mathematical truth}" (\emph{ibid}, pp. 76).

In `Shadows of the Mind' Penrose puts forward two distinct lines of argument; a broad argument and a more nuanced one:
\begin{itemize}
\item The `broad' argument is essentially the `core' Penrose-Lucas position (in the context of mathematicians' belief that they really are ``doing what they think they are doing'', contra blindly following the rules of an unfathomably complex algorithm), such that ``the procedures available to the mathematicians ought all to be knowable''. This argument leads Penrose to conclusion \textbf{G} (above).
\item More nuanced lines of argument, addressed at those who take the view that mathematicians are not ``really doing what they think they are doing'', but are merely acting like Searle in the Chinese room and blindly following the rules of a complex, unfathomable rule-book. In this case, as there is no way to know what the algorithm is, Penrose instead examines how it might conceivably have come about, considering (a) the role of natural selection and (b) some form of engineered construction (e.g. neural network, evolutionary computing, machine learning etc); a discussion of these lines of argument is outside the scope of this paper. 
\end{itemize}

\subsection{The basic Penrose' argument}
Consider $a$ to be a `\emph{knowably sound}' sound set of rules (an effective procedure) to determine if $C (n)$ - the computation $C$ on the natural number $n$ (e.g. `\emph{Find an odd number that is the sum of $n$ even numbers}') - does not stop. Let $A$ be a formalisation of all such effective procedures known to human mathematicians. By definition, the application of $A$ terminates iff $C (n)$ does not stop. Now, consider a human mathematician continuously analysing $C (n)$ using the effective procedures, $A$, and only halting analysis if it is established that $C (n)$ does not stop.

NB. $A$ must be `\emph{knowably sound}' and cannot be wrong if it decides that $C (n)$ does not stop because, Penrose claims, if $A$ was `knowably sound' and if any of the procedures in $A$ were wrong, the error would eventually be discovered.

Computations of one parameter, $n$, can be enumerated (listed):\\
\begin{center}
C$_0 (n), C_1 (n), C_2 (n) .. C_p (n)$\\
\end{center}
where $C_p (n)$ is the $p^{th}$ computation on $n$ (i.e. it defines the $p^{th}$ computation of one parameter $n$). Hence $A (p, n)$ is the effective procedure that, when presented with $p$ and $n$, attempts to discover if $C_p (n)$ will not halt. I.e. If $A (p, n)$ ever halts, then we know for certain that $C_p (n)$ does not halt\footnote{Penrose, `Shadows of the Mind' (pp. 72-77).}.

Given the above, Penrose' simple G\"{o}delian argument can be summarised as follows: 

\begin{quote}
\begin{qFont}
\begin{enumerate}
\item \label{rp1} If $A (p, n)$ halts then $C_p (n)$ does not halt.

\item \label{rp2} Now consider the `Self-Applicability Problem' (SAP), by letting $p = n$ in statement (\ref{rp1}) above; thus:

\item \label{rp3} If $A (n, n)$ halts then $C_n (n)$ does not halt.

\item \label{rp4} But $A (n, n)$ is a function of one natural number, $n$ and hence must be found in the enumeration of $C$. Let us assume it is found at position $k$ (i.e. it is the $k_{th}$ computation of one parameter $C_k (n)$); thus:

\item \label{rp5} $A (n, n) = C_k (n)$. 

\item \label{rp6} \emph{Now, consider the particular computation where $n = k$}; i.e. substituting $n = k$ into statement (\ref{rp5}) above; thus:

\item \label{rp7} $A (k, k) = C_k (k)$.

\item \label{rp8}And rewriting (\ref{rp3}) with $n = k$; thus: 

\item \label{rp9} iff $A (k, k)$ halts then $C_k (k)$ does not halt.

\item \label{rp10}But substituting from (\ref{rp7}) into (\ref{rp9}), we get the following; thus:

\item \label{rp11} If $C_k (k)$ halts then $C_k (k)$ does not halt, which clearly leads to contradiction \textbf{if $C_k (k)$ halts}.
 
\item \label{rp12} Hence from (\ref{rp11}) we know that, if $A$ is sound (and there is no contradiction) \textbf{then $C_k (k)$ cannot halt}. 

\item \label{rp13} However, $A$ cannot itself signal (\ref{rp12}) [by halting] because (\ref{rp7}): $A (k, k) = C_k (k)$. I.e. if $C_k (k)$ cannot halt then $A (k, k)$ cannot either.

\item \label{rp14} Furthermore, if $A$ exists \textbf{and is sound} then \textbf{we know} $C_k (k)$ cannot halt; however $A$ is provably incapable of ascertaining this, because we also know (from statement (\ref{rp11})) that $A$ halting [to signal that $C_k (k)$ cannot halt] would lead to contradiction.

\item \label{rp15} So, if $A$ exists and is sound, we \textbf{know} (from statement (\ref{rp12})) that $C_k (k)$ cannot halt, and hence we know something (via statement (\ref{rp13})) that $A$ is provably unable to ascertain (\ref{rp14}).

\item \label{rp16} Hence $A$ - \emph{the \textbf{formalisation} of all procedures known to mathematicians} - cannot encapsulate human mathematical understanding.
\end{enumerate}
\end{qFont}
\end{quote}

In other words, the human mathematician can `see' that the G\"{o}del Sentence is true for consistent $F$, even though the consistent $F$ cannot prove $G(\check{g})$. 

Arguments targeting computationalism on the basis of G\"{o}delian theory have been vociferously critiqued ever since they were first made\footnote{Lucas maintains a web page \url{http://users.ox.ac.uk/\~{}jrlucas/Godel/referenc.html} listing over fifty such criticisms; see also \citep{Psyche_1995} for extended peer commentary specific to the Penrose version.}, however discussion - both negative and positive - still continues to surface in the literature\footnote{Cf. \citep{Bringsjord_2000} and \citep{Tassinari_2007}.} and detailed review of their absolute merit falls outside the scope of this work. In this context it is sufficient simply to note, as the philosopher John Burgess wryly observed, that the Penrose-Lucas thesis may be fallacious but ``\emph{logicians are not unanimously agreed as to where precisely the fallacy in their argument lies}''  \citep{Burgess_2000}. Indeed Penrose, in response to a volume of peer commentary on his argument \citep{Psyche_1995}, ``\emph{was struck by the fact that none of the present commentators has chosen to dispute my conclusion} \textbf{G}:''  \citep{Penrose_1996}. 

Perhaps reflecting this, after a decade of robust international debate on these ideas, in 2006 Penrose was honoured with an invitation to present the opening public address at `Horizons of truth', the G\"{o}del centenary conference at the University of Vienna; for Penrose, G\"{o}delian arguments continue to suggest human consciousness cannot be realised by algorithm; there must be a ``\emph{noncomputational ingredient in human conscious thinking}'' \citep{Penrose_1996}.

\section{Consciousness, computation and panpsychism}
\begin{figure}
\begin{center}
\includegraphics [scale=0.4]{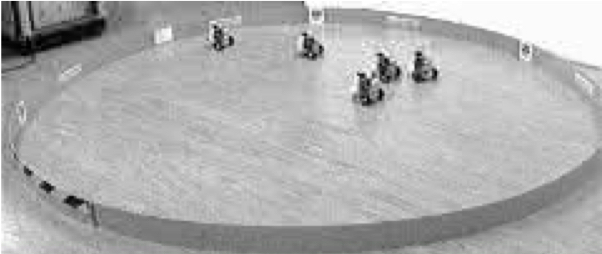}
\caption{Kevin Warwick's `Seven Dwarves': neural network controlled robots.}
\label{7Dwarves}
\end{center}
\end{figure}

Figure (\ref{7Dwarves}) shows Professor Kevin Warwick's ``Seven Dwarves'' cybernetic learning robots in the act of moving around a small coral, `learning' not to bump into each other. Given that (i) in `learning', the robots developed individual behaviours and (ii) their neural network controllers used approximately the same number of `neurons' as found in the brain of a slug, Warwick has regularly delighted in controversially asserting that the robots were ``\emph{as conscious as a slug}'' and that it is only ``\emph{human bias}'' (human chauvinism) that has stopped people from realising and accepting this \citep{Warwick_2002}. Conversely, even as a fellow cybernetician and computer scientist, I have always found such remarks - that the mechanical execution of appropriate computation [by a robot] will realise consciousness - a little bizarre, and eventually derived the following, a priori, argument to highlight the implicit absurdness of such claims.

The Dancing with Pixies (DwP) \emph{reductio ad absurdum} \citep{Bishop_2002a} is my attempt to target any claim that machines (qua computation) can give rise to raw sensation (phenomenal experience), unless we buy into a very strange form of panpsychic mysterianism. Slightly more formally, DwP is a simple \emph{reductio ad absurdum} argument to demonstrate that \emph{if} ([appropriate] computations realise phenomenal sensation in machine) \emph{then} (panpsychism holds). I.e. \emph{If} the DwP is correct \emph{then} we must either accept a vicious form of panpsychism (wherein every open physical system is phenomenally conscious) \emph{or} reject the assumed claim (computational accounts of phenomenal consciousness). Hence, because panpsychism has come to seem an implausible world view\footnote{Framed by the context of our immense scientific knowledge of the closed physical world, and the corresponding widespread desire to explain everything ultimately in physical terms.}, we are obliged to reject any computational account of phenomenal consciousness.

At its foundation, the core DwP reductio (\emph{ibid}) derives from an argument by Hilary Putnam, first presented in the Appendix to `Representation and Reality' \citet{Putnam_1988}; however, it is also informed by \citet{Maudlin_1989} (on computational counterfactuals), \citet{Searle_1990} (on software isomorphisms) and subsequent criticism from \citet{Chalmers_1996}, \citet{Klein_2018} and \citet{Chrisley_1995}\footnote{For early discussion of these themes see `Minds and Machines', \textbf{4: 4}, `What is Computation?', November 1994.}. Subsequently, the core DwP argument has been refined, and responses to various criticisms of it presented, across a series of papers \citet{Bishop_2002b, Bishop_2002a, Bishop_2009b, Bishop_2014}. For the purpose of this review, however, I merely present the heart of the reductio.

In the following discussion, instead of seeking to justify the claim from \citep{Putnam_1988}, that ``\emph{every ordinary open system is a realization of every abstract finite automaton}'' (and hence that, ``\emph{psychological states of the brain cannot be functional states of a computer}''), I will show that, over any finite time period, every open physical system implements the particular execution trace [of state transitions] of a computational system $Q$, operating on known input $I$. That this result leads to panpsychism is clear as, equating $Q (I)$ to a specific computational system (that is claimed to instantiate phenomenal experience as it executes), and following Putnam's state-mapping procedure, an identical execution trace of state transitions (and \emph{ex hypothesi} phenomenal experience) can be realised in any open physical system.

\subsection{The Dancing with Pixies (DwP) reductio ad absurdum}
\begin{figure}
\begin{center}
\includegraphics [scale=0.12]{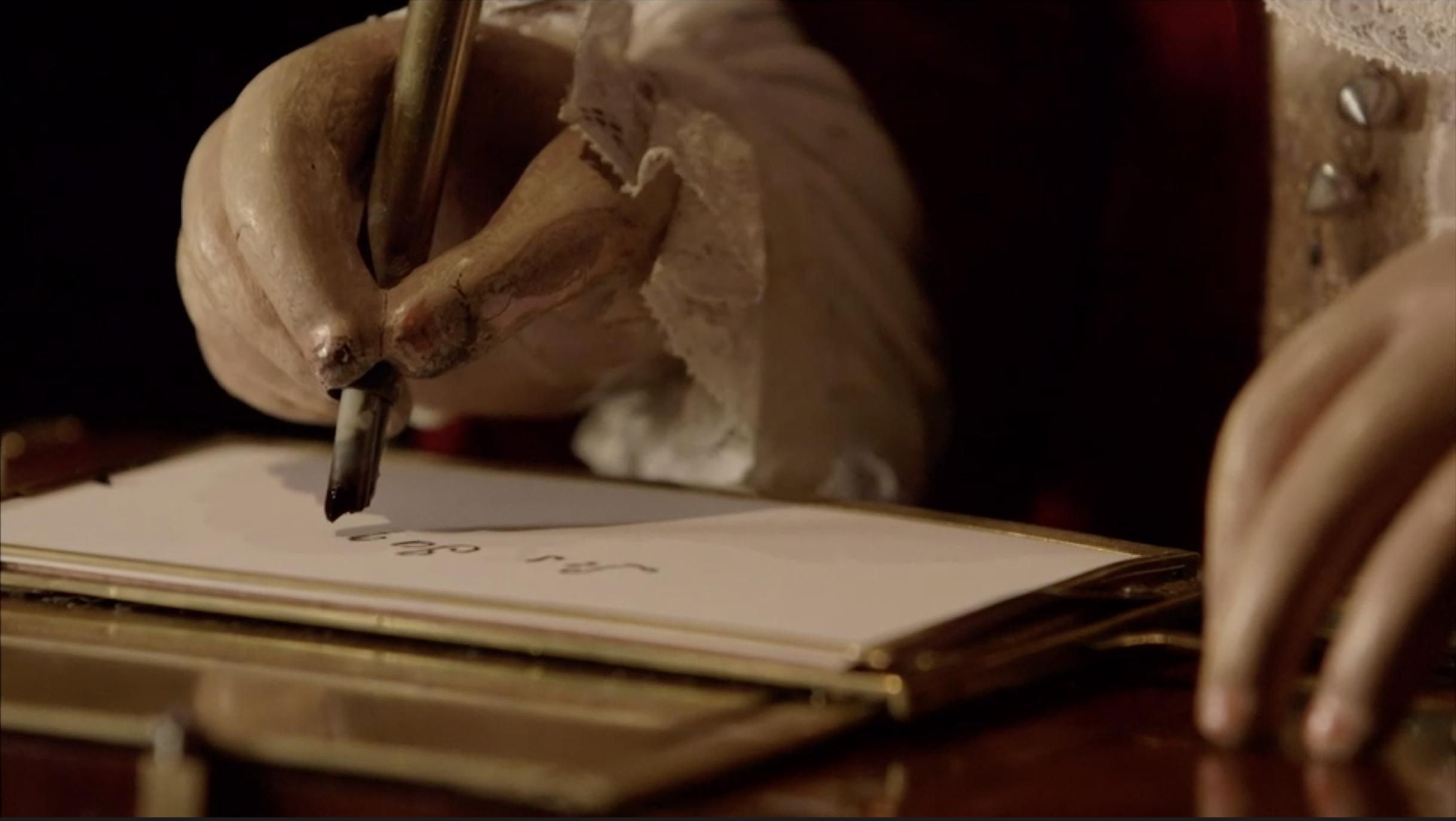}
\caption{Photograph of Jaquet Droz' The Writer (image screenshot from BBC4 `Mechanical Marvels Clockwork Dreams: The Writer [2013]).}
\label{TheWriter}
\end{center}
\end{figure}

Perhaps you have seen an automaton at a museum or on television; `The Writer' is one of three surviving automata from the 18th century built by Jaquet Droz and was the inspiration for the movie Hugo; it still writes today (see Figure (\ref{TheWriter})). The complex clockwork mechanism seemingly brings the automaton to life as it pens short (`pre-programmed') phrases. Such machines were engineered to follow through a complex sequence of operations - \emph{in this case, to write a particular phrase} - and to early-eyes at least, and even though they are insensitive to real-time interactions, appeared almost sentient; uncannily\footnote{Sigmund Freud first introduced the concept of `the uncanny' in his 1919 essay `Das Unheimliche' \citep{Freud_1919}, which explores the eeriness of dolls and waxworks; subsequently, in aesthetics, `the uncanny' highlights a hypothesized relationship between the degree of an object's resemblance to a human being and the human emotional response to such an object. The notion of the `uncanny' predicts humanoid objects which imperfectly resemble real humans, may provoke eery feelings of revulsion and dread in observers \citep{MacDorman_2006}. \citep{Mori_2012} subsequently explored this concept in robotics, through the notion of `the uncanny valley'. Recently, the notion of the uncanny has  been critically explored through the lens of feminist theory and contemporary art practice, for example by Alexandra Kokoli who, in focussing on Lorraine O'Grady performances as a ``black feminist killjoy'', stridently calls out ``the whiteness and sexism of the artworld'' \citep{Kokoli_2016}.} life-like in their movements. 

In his 1950 paper Computing Machinery and Intelligence \citet{Turing_1950} described the behaviour of a simple physical automaton - his `Discrete State Machine'. This was a simple device with one moving arm, like the hour hand of a clock; with each tick of the clock Turing conceived the machine cycling through the 12 o'clock, 8 o'clock and 4 o'clock positions. Turing (\emph{ibid}) showed how we can describe the state evolution of his machine as a simple Finite State Automaton (FSA).

\begin{figure}
\begin{center}
\includegraphics [scale=0.15]{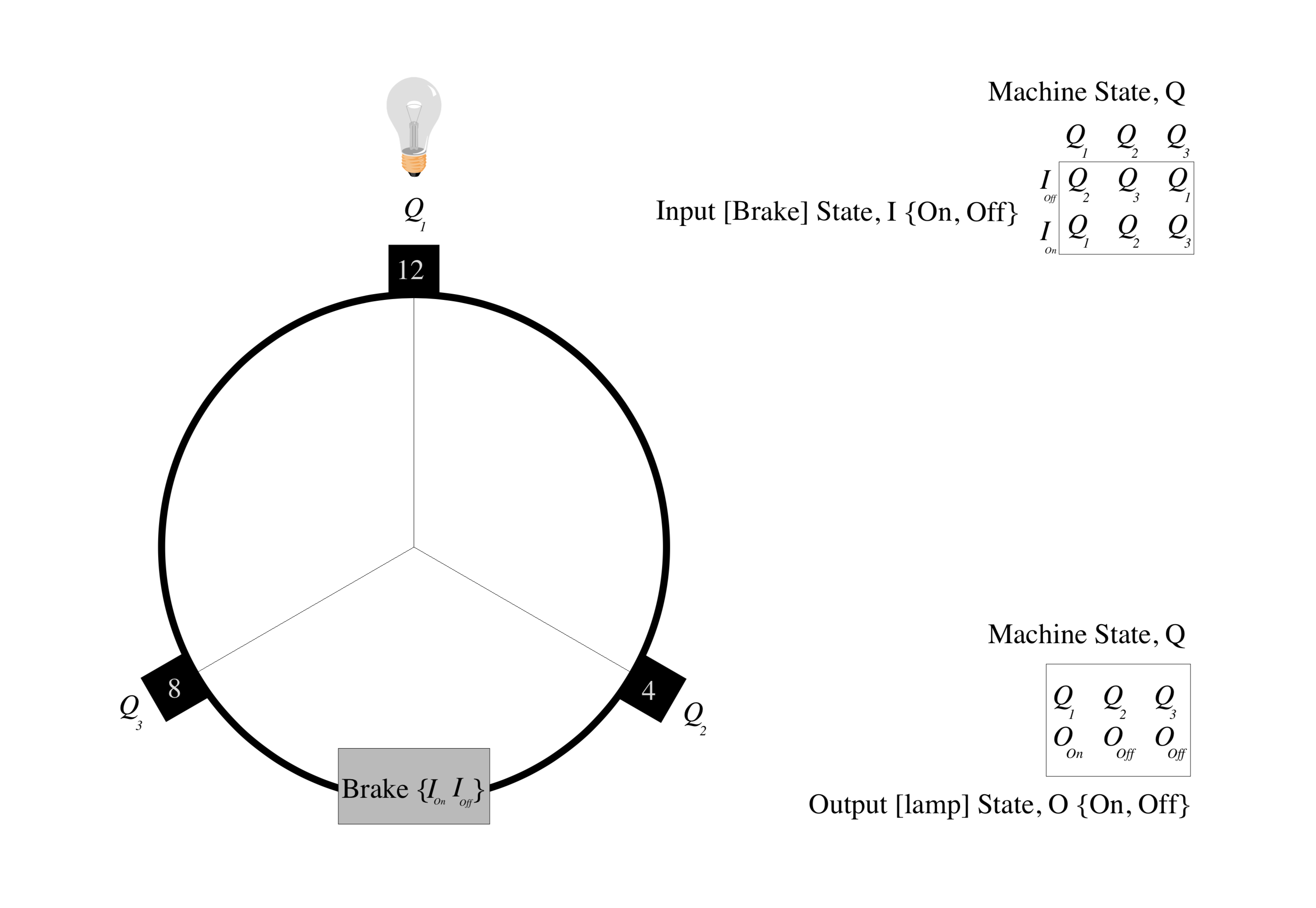}
\caption{Turing's Discrete State Machine.}
\label{TuringDSM}
\end{center}
\end{figure}

Turing assigned the 12 o'clock (noon/midnight) arm position to FSA state (machine-state) $Q_1$; the 4 o'clock arm position to FSA state $Q_2$ and the 8 o'clock arm position to FSA state $Q_3$. N.B. Turing's mapping of the machine's physical arm position to a logical FSA (computational) state is arbitrary (e.g. Turing could have chosen to assign the 4 o'clock arm position to FSA state $Q_1$)\footnote{In any electronic digital circuit, it is an engineering decision, contingent on the type of logic used - TTL, ECL, CMOS etc. - what voltage range corresponds to a logical TRUE value and what range to a logical FALSE.}. The machine's behaviour can now be described by a simple \emph{state-transition table}: if the FSA is in state $Q_1$ then it goes to FSA state $Q_2$; if in FSA state $Q_2$ it goes to $Q_3$; if in FSA state $Q_3$ goes to $Q_1$. Hence, with each clock tick the machine will cycle through FSA states $Q_{1}, Q_{2}, Q_{3}, Q_{1}, Q_{2}, Q_{3}, Q_{1}, Q_{2}, Q_{3}, ...$ etc., (as shown in Figure (\ref{TuringDSM})).

To see how Turing's machine could control Jaquet Droz' Writer automaton, we simply need to ensure that when the FSA is in a particular machine-state, a given action is caused to occur. For example, if the FSA is in FSA state $Q_{1}$ then, say, a light might be made to come on, or The Writer's pen be moved. In this way complex sequences of actions can be `programmed'.

Now, what is perhaps not so obvious is that, over any given time-period, we can fully emulate Turing's machine with a simple digital counter (e.g. a digital milometer); all we need to do is to \emph{map} the digital counter state $C$ to the appropriate FSA state $Q$. I.e, If the counter is in state $C_0 =$ \{$000000$\} then we map to FSA state $Q_{1}$; if it is $C_1 =$ \{$000001$\} then we map to FSA state $Q_{2}$; \{$000002$\} $\rightarrow Q_{3}$; \{$000003$\} $\rightarrow Q_{1}$; \{$000004$\} $\rightarrow Q_{2}$; \{$000005$\} $\rightarrow Q_{3}$, ... etc. 

Thus, if the counter is initially in state $C_0 =$ \{$000000$\} then, over the time interval [$t=0 .. t=5$], it will reliably transit states \{$000000 \rightarrow 000001 \rightarrow 000002 \rightarrow 000003 \rightarrow 000004 \rightarrow 000005$\} which, by applying the Putnam mapping defined above, generates the Turing FSA state sequence: \{$Q_{1} \rightarrow Q_{2} \rightarrow Q_{3} \rightarrow Q_{1} \rightarrow Q_{2} \rightarrow Q_{3}$\} over the interval  [$t=0 .. t=5$]. In this manner any input-less FSA can be realised by a [suitably large] digital counter. 

Furthermore, sensu stricto, all \emph{real} computers (machines with finite storage) are Finite State Machines\footnote{Even if we usually think about computation in terms of the [more powerful] Turing Machine model.} and so a similar process can be applied to any computation realised by a PC. However, before looking to replace your desktop machine with a simple digital counter, keep in mind that a FSA without input is an extremely trivial device (as is evidenced by the ease in which it can be emulated by a simple digital counter), merely capable of generating a single unbranching sequence of states ending in a cycle, or at best in a finite number of such sequences (e.g. \{$Q_{1} \rightarrow Q_{2} \rightarrow Q_{3} \rightarrow Q_{1} \rightarrow Q_{2} \rightarrow Q_{3}$\}, ... etc.).

However, Turing also described the operation of a discrete state machine with input in the form of a simple lever-brake mechanism, which could be made to either lock-on (or lock-off) at each clock-tick. Now, if the machine is in computational state \{$Q_1$\} and the brake is on, then the machine stays in  \{$Q_1$\} otherwise it moves to computational state  \{$Q_2$\}; if machine is in \{$Q_2$\} and brake is on, it stays in \{$Q_2$\} otherwise it goes to \{$Q_3$\} and if machine is in state \{$Q_3$\} and brake is on, it stays in \{$Q_3$\} otherwise it cycles back to state \{$Q_1$\}. In this manner, the addition of input has transformed the machine, from a simple device that could merely cycle through a simple unchanging list of states, to one that is sensitive to input and as a result the number of possible state sequences that it may enter grows combinatorially with time, rapidly becoming larger than the number of atoms in the known universe. It is due to this exponential growth in potential state transition sequences that we cannot, so easily, realise a FSA with input (or a PC) using a simple digital counter. 

Nonetheless, if we have \emph{knowledge} of the input over a given time period (say, we \emph{know} that the brake is initially ON for the first clock tick and OFF thereafter), then the combinatorial contingent state structure of an FSA with input, simply collapses into a simple linear list of state transitions (e.g. \{$Q_{1} \rightarrow Q_{2} \rightarrow Q_{3} \rightarrow Q_{1} \rightarrow Q_{2} \rightarrow Q_{3}$\}, ... etc.), and so once again can be simply realised by a suitably large digital counter using the appropriate Putnam mapping. 

Thus, to realise Turing's machine, say, with the brake ON for the first clock tick and OFF thereafter, we simply need to specify that the initial counter in state \{$000000$\} maps to the first FSA state $Q_1$; state \{$000001$\} maps to FSA state $Q_1$; \{$000002$\} maps to $Q_2$; \{$000003$\} to $Q_3$; \{$000004$\} to $Q_1$; \{$000005$\} to $Q_2$ etc.).

In this manner, considering the execution of any putative machine consciousness software that is claimed to be conscious (e.g. the control program of Kevin Warwick's robots) if, over a finite time period, we know the input\footnote{E.g. We can obtain the input to a robot [that is claimed to experience phenomenal consciousness as it interacts with the world] by deploying a `data-logger' to record the data obtained from all its various sensors etc.}, we can generate precisely the same state transition trace with any [suitably large] digital counter. Furthermore, as Hilary Putnam demonstrated, in place of using a digital counter to generate the state sequence \{$C$\}, we could deploy \emph{any} `open physical system' (such as a rock\footnote{The `Principle of Noncyclical Behaviour', \citep{Putnam_1988}, asserts: a system $S$ is in different `maximal states' \{$S_1, S_2, S_n$\} at different times. This principle will hold true of all systems that can ``see'' (are not shielded from electromagnetic and gravitational signals from) a clock. Since there are natural clocks from which no ordinary Open system is shielded, all such systems satisfy this principle. (N.B.: It is not assumed that this principle has the status of a physical law; it is simply assumed that it is in fact true of all ordinary macroscopic open systems).}) to generate a suitable non-repeating state sequence \{$S_1, S_2, S_3, S_4, ... $\}, and map FSA states to these [non-repeating] `rock' states \{$S$\} instead of the counter states. Following this procedure a rock, alongside a suitable Putnam mapping, can made to realise any finite series of state transitions.

Thus, if any AI system is phenomenally conscious\footnote{E.g. Perhaps it `sees' the ineffable red of a rose; smells its bouquet etc.} as it executes a specific set of state transitions over a finite time period, then a vicious form of panpsychism must hold, because the same raw sensation, phenomenal consciousness, could be realised with a simple digital counter (a rock, or \emph{any open physical system}) and the appropriate Putnam mapping. In other words, unless we are content to `bite the bullet' of panpsychism, then no machine, however complex, can ever realise phenomenal consciousness purely in virtue of the execution of a particular computer program\footnote{In \citep{Bishop_2017}, I consider the further implications of the DwP reductio for `digital ontology' and the Sci-Fi notion, pace \citep{Bostrom_2003}, that we are `most likely' living in a digitally simulated universe.}.

\section{Conclusion} 
It is my contention that at the heart of classical cognitive science - artificial neural networks, causal cognition and artificial intelligence - lies a ubiquitous computational metaphor:
\begin{itemize}
\item \textbf{Explicit computation}: cognition as `computations on symbols'; GOFAI; [physical] symbol systems; functionalism (philosophy of mind); cognitivism (psychology); language of thought (philosophy; linguistics).
\item \textbf{Implicit computation}: cognition as `computations on sub-symbols'; connectionism (sub-symbolic AI; psychology; linguistics); the digital connectionist theory of mind (philosophy of mind).
\item \textbf{Descriptive computation}: neuroscience as `computational simulation'; Hodgkin-Huxley mathematical models of neuron action potentials (computational neuroscience; computational psychology).
\end{itemize}

In contrast, the three arguments outlined in this paper purport to demonstrate: (i) that computation cannot realise understanding; (ii) that computation cannot realise mathematical insight and (iii) that computation cannot realise raw sensation, and hence that computational syntax will never fully encapsulate human semantics. Furthermore, these a priori arguments pertain to all possible computational systems, whether they be driven by `Neural Networks\footnote{Including `Whole Brain Emulation' and, a fortiori, Henry Markram's `Whole Brain Simulation', as underpins both the `Blue Brain Project' - \emph{a Swiss research initiative that aimed to create a digital reconstruction of rodent and eventually human brains by reverse-engineering mammalian brain circuitry} - and the concomitant, controversial, EUR 1.019 billion flagship European `Human Brain Project' \citep{Fan_2019}.}', `Bayesian Networks' or a `Causal Reasoning' approach. 

Of course, `deep understanding' is not always required to engineer a device to do $x$, but when we do attribute agency to machines, or engage in unconstrained, unfolding interactions with them, `deep [human-level] understanding' matters. In this context, it is perhaps telling that after initial quick gains in the average length of interactions with her users, XiaoIce has been consistently performing no better than, on average, 23 conversational turns for a number of years now\footnote{Although it is true to say than many human-human conversations don't even last this long - a brief exchange with the person at the till in a supermarket -  in principle, with sufficient desire and shared interests, human conversations can be delightfully open ended.}. Although chatbots like XiaoIce and Tay will continue to improve, lacking genuine understanding of the bits they so adroitly manipulate, they will ever remain prey to egregious behaviour of the sort that finally brought Tay offline in March 2016, with potentially disastrous brand consequences\footnote{Cf. Tay's association with `racist' tweets or Apple's association with `allegations of gender bias' in assessing applications for its credit card \url{https://www.bbc.co.uk/news/business-50432634}.}.

Techniques such as `causal cognition' - which focuses on mapping and understanding the cognitive processes that are involved in perceiving and reasoning about cause-effect relations - whilst undoubtedly constituting a huge advance in the mathematization of causation will, on its own, move us no nearer to solving foundational issues in AI pertaining to teleology and meaning. Whilst causal cognition will undoubtedly be helpful in engineering specific solutions to particular human specified tasks, lacking human understanding, the dream of creating an AGI remains as far away as ever. Without genuine understanding, the ability to seamlessly transfer \emph{relevant} knowledge from one domain to another will remain allusive. Furthermore, lacking phenomenal sensation (in which to both ground meaning and desire), even a system with a 'complete explanatory model' (allowing it to accurately predict future states) would still lack intentional \emph{pull}, with which to drive genuinely autonomous teleological behaviour\footnote{Cf. Raymond Tallis, \emph{How On Earth Can We Be Free?} \url{https://philosophynow.org/issues/110/How_On_Earth_Can_We_Be_Free}.}.

No matter how sophisticated the computation is, how fast the CPU is or how great the storage of the computing machine is, there remains an unbridgeable gap (a `humanity gap') between the engineered problem solving ability of machine and the general problem solving ability of man\footnote{Within cognitive science there is an exciting new direction broadly defined by the so-called 4Es: the Embodied, Enactive, Ecological and Embedded approaches to cognition (cf. \citep{Thompson_2007}); together, these offer an alternative approach to meaning, grounded in the body and environment, but at the cost of fundamentally moving away from the computationalist's vision of the multiple realisability [in silico] of cognitive states.}. As a source close to the autonomous driving company Waymo\footnote{An American autonomous driving technology development company; a subsidiary of Alphabet Inc, the parent company of Google.} recently observed (in the context of autonomous vehicles): 

\begin{quote}
\begin{qFont}
``There are times when it seems autonomy is around the corner and the vehicle can go for a day without a human driver intervening ... other days reality sets in because \textbf{the edge cases are endless} ...'' (The Information: 28th August, 2018).
\end{qFont}
\end{quote}



\newpage

\bibliography{mybib}

\end{document}